\documentstyle[aps]{revtex} 
\begin{document}
\newcommand{\s}{\varsigma}
\newcommand{\room}{\rule[-0.3cm]{0cm}{0.8cm}}
\newcommand{\hsp}{\hspace*{3mm}}
\newcommand{\vsp}{\vspace*{3mm}}
\newcommand{\nsp}{\vspace*{-3mm}}
\newcommand{\be}{\begin{equation}}
\newcommand{\ee}{\end{equation}}
\newcommand{\bd}{\begin{displaymath}}
\newcommand{\ed}{\end{displaymath}}
\newcommand{\bdm}{\begin{displaymath}}
\newcommand{\edm}{\end{displaymath}}
\newcommand{\bea}{\begin{eqnarray}}
\newcommand{\eea}{\end{eqnarray}}
\newcommand{\sgn}{~{\rm 
sgn}}
\newcommand{\extr}{~{\rm extr}}
\newcommand{\Equiv}{\Longleftrightarrow}
\newcommand{\pprime}{\prime\prime}
\newcommand{\notexists}{\exists\hspace*{-2mm}/}
\newcommand{\bra}{\langle}
\newcommand{\ket}{\rangle}
\newcommand{\bigbra}{\left\langle\room}
\newcommand{\bigket}{\right\rangle\room}
\newcommand{\bras}{\langle\!\langle}
\newcommand{\kets}{\rangle\!\rangle_{xy}}
\newcommand{\bigbras}{\left\langle\!\!\!\left\langle\room}
\newcommand{\bigkets}{\right\rangle\!\!\!\right\rangle_{\!\!xy}\room}
\newcommand{\order}{{\cal 
O}}
\newcommand{\minus}{\!-\!}
\newcommand{\plus}{\!+\!}
\newcommand{\erf}{{\rm 
erf}}
\newcommand{\bk}{\mbox{\boldmath $k$}}
\newcommand{\bm}{\mbox{\boldmath 
$m$}}
\newcommand{\br}{\mbox{\boldmath $r$}}
\newcommand{\bq}{\mbox{\boldmath 
$q$}}
\newcommand{\bz}{\mbox{\boldmath $z$}}
\newcommand{\bH}{\mbox{\boldmath 
$H$}}
\newcommand{\bM}{\mbox{\boldmath $M$}}
\newcommand{\bQ}{\mbox{\boldmath 
$Q$}}
\newcommand{\bR}{\mbox{\boldmath $R$}}
\newcommand{\bS}{\mbox{\boldmath 
$S$}}
\newcommand{\calD}{{\cal D}}
\newcommand{\calF}{{\cal F}}
\newcommand{\calP}{{\cal 
P}}
\newcommand{\calS}{{\cal S}}
\newcommand{\bW}{\mbox{\boldmath $W$}}
\newcommand{\hbh}{\hat{\mbox{\boldmath 
$h$}}}
\newcommand{\hbm}{\hat{\mbox{\boldmath $m$}}}
\newcommand{\hbr}{\hat{\mbox{\boldmath 
$r$}}}
\newcommand{\hbq}{\hat{\mbox{\boldmath $q$}}}
\newcommand{\hbD}{\hat{\mbox{\boldmath 
$D$}}}
\newcommand{\hbQ}{\hat{\mbox{\boldmath $Q$}}}
\newcommand{\hbR}{\hat{\mbox{\boldmath 
$R$}}}
\newcommand{\hbW}{\hat{\mbox{\boldmath $W$}}}
\newcommand{\bsigma}{\mbox{\boldmath 
$\sigma$}}
\newcommand{\bomega}{\mbox{\boldmath $\Omega$}}
\newcommand{\bphi}{\mbox{\boldmath 
$\Phi$}}
\newcommand{\bpsi}{\mbox{\boldmath $\psi$}}
\newcommand{\bdelta}{\mbox{\boldmath 
$\Delta$}}
\newcommand{\btheta}{\mbox{\boldmath $\theta$}}
\newcommand{\bxi}{\mbox{\boldmath 
$\xi$}}
\newcommand{\bmu}{\mbox{\boldmath $\mu$}}
\newcommand{\brho}{\mbox{\boldmath 
$\rho$}}
\newcommand{\bEta}{\mbox{\boldmath $\eta$}}
\newcommand{\req}{r_{\rm 
eq}}
\newcommand{\unity}{{\bf 1}\hspace{-1mm}{\bf I}}

\draft

\title{
\centerline{\bf
Macrodynamics 
of Disordered and Frustrated Systems}
}

\author{\bf
{D. Sherrington} \rm(1,2)\, 
\bf{A.C.C. Coolen} \rm(2,3)\ and 
\bf{S.N. Laughton} \rm(2)\}}
\address{(1) 
Center for Nonlinear Studies, Los Alamos National
Laboratory 
\linebreak 
(2) Physics Department, University of Oxford,
 England \linebreak (3) Mathematics 
Department, Kings College, London,
England}
\maketitle
 
\begin{center} 
\today \end{center}

\begin{abstract} 
It is shown how the macroscopic 
non-equilibrium dynamics of a class of 
systems whose microscopic stochastic 
dynamics involves disordered and 
frustrated but range-free interactions 
can be well described by closed
deterministic flow equations; this requires 
an appropriate choice of
order parameters/function and ans\"atze. 
\end{abstract}

\section{Introduction}

Given 
a system of many macroscopic units driving one another
dynamically 
through 
strongly interactive, possibly mutually competitive and 
stochastically 
applied rules, one is often interested in determining 
the temporal development 
of some macroscopic observable(s). In general
this is a difficult task, 
normally impossible without approximation. 
However, when (i) the individual 
units are identical, (ii) the
microscopic 
update rules are instantaneous, 
and (iii) there is no spatial
dependency 
to the interaction rules, experience 
with extremal dominance in 
equilibrium physics suggests that solution 
in a closed form for an 
appropriate set of observables (or order parameters) 
becomes at least 
potentially feasible in the limit in which the number 
of units $N$
tends 
to infinity. Indeed, for simple enough units with uniform 
interaction 
rules, a closed autonomous dynamical equation can be obtained
involving 
only a single order parameter. On the other hand, when the 
interaction 
rules are non-uniform and involve significant random
character 
the situation 
is very non-trivial even for the case of range-free 
interactions. Nevertheless, 
we have been able to make major progress 
with such problems, using a combination 
of exact results and ans\"atze
to obtain closed autonomous macroscopic 
flow equations whose solutions
are in very good accord with the results 
of numerical simulations of
the 
microdynamics and suggest procedures to 
tackle other analagous and
extended 
problems. In this paper we spell out 
the philosophy and execution of 
this new approach and demonstrate its 
application to simple models of
spin 
glasses and recurrent neural networks.

\section{The problem}

In general terms, we are concerned with systems 
whose microscopic
state is 
described by a set of $N$ variables ${S}_i$; 
$i=1...N$ which obey
random 
stochastic microdynamics based on the instantaneous 
state of the other
spins. 
Our objective is to devise a description of 
the subsequent
macrodynamics in 
terms of closed equations for fewer-parameter 
sets of 
macrovariables $\Omega_\mu(\bS);\mu=1...n$.

More specifically, 
let us initially concentrate on systems in which
the 
variables $\bS$ are 
binary spins $\{\sigma_i=\pm 1\}$ and obey random 
sequential Glauber stochastic 
dynamics via local effective fields
determined 
through pairwise exchange 
interactions with other spins and external
stimuli. 
Thus, the spins are 
updated randomly sequentially with transition 
probabilities
\begin{equation}
p(\sigma_i 
\rightarrow \sigma'_i) = \frac{1}{2} \left[1+\sigma'_i
\tanh[\beta h_i(\bsigma)]\right]
\label{eq:Glauber}
\end{equation}
where
\begin{equation}
h_{i}(\bsigma) 
= \sum_{j\neq i}J_{ij}\sigma_{j}+\theta_{i}
\label{eq:field}
\end{equation}
and 
$\beta^{-1}$ is a measure of the degree of stochasticity. With 
appropriate 
coarse-graining this leads to the master equation for the 
microstate distribution 
$p_t({\bsigma})$
\begin{equation}
\frac{d}{dt}p_{t}({\bsigma})=\sum_{k=1}^{N} 
\left[p_{t}(F_{k}\bsigma)
W_{k}(F_{k}\bsigma) - p_{t}(\bsigma)W_{k}(\bsigma)\right]
\label{eq:master}
\ee
where 
$F_k$ is the spin-flip operator 
\begin{equation} 
F_k\Phi(\bsigma) = 
\Phi (\sigma_i,...,-\sigma_k,...,\sigma_N), 
\label{eq:spinflip}
\end{equation}
$W_k(\bsigma)$ 
is the transition rate 
\be
W_k(\bsigma) = {1\over 2}[1-\sigma_k\tanh(\beta 
h_k(\bsigma))],  
\label{eq:transition}
\ee
and we are now using the vector 
notation $\bsigma\ = (\sigma_1,...,\sigma_N)$.

From (\ref{eq:master}) 
we may derive an equation for the evolution of
the 
macrovariable probability 
distribution 

\begin{equation}
P_t[\bomega]=\sum_{\bsigma} p_t (\bsigma)
\delta[\bomega-\bomega(\bsigma)]; 
\ \ \  \bomega\equiv(\Omega_1,...\Omega_n)
\label{eq:Pomega}
\end{equation}
in 
the form
\begin{equation}
\frac{d}{dt} P_t[\bomega]=
\sum_{\ell\geq 1} 
\frac{(-1)}{\ell !}^\ell \sum^n_{k_{1}=1}.. 
\sum^n_{k_{\ell}=1} 
\frac{\partial^\ell}{\partial\Omega_{k_1}...\partial\Omega_{k_{\ell}}}
P_t[\bomega]F^{(\ell)}_{k_{1}..k_{\ell}} [\bomega ;t]
\label{eq:dPomegadt}
\end{equation}
where
\begin{equation}
F^{(\ell)}_{k_{1}...k_{\ell}}[\bomega
;t]=\langle\sum^N_{j=1}W_j(\bsigma)
\Delta_{jk_{1}}(\bsigma)...\Delta_{jk_{\ell}}(\bsigma)\rangle_{\bomega;t}
\ 
\ \ \ \ \Delta_{jk}(\bsigma)\equiv\Omega_k(F_j\bsigma)-\Omega_k(\bsigma)
\label{eq:F}
\end{equation}
and 
the notation $\langle\rangle_{\bf{\Omega};t}$ refers to a sub-shell average
\begin{equation}
\langle 
f(\bsigma)\rangle _{\bomega;t}\equiv
\frac{\sum_{\bsigma}p_t({\bsigma})
\delta[{\bomega}-{\bomega}({\bsigma})]f({\bsigma})}{\sum_{\bsigma}p_t({\bsigma})\delta[{\bomega}-{\bomega}({\bsigma)]}} 
\ \ .
\label{eq:f}
\end{equation}

In several cases of interest and for 
finite times only the first term
on the 
right hand side of (\ref{eq:dPomegadt}) 
survives in the limit
$N\to\infty$, 
yielding the deterministic flow
\begin{equation}
\frac{d}{dt}{\bomega}_t 
=
\langle\sum_iW_i({\bsigma})[{\bomega}(F_i{\bsigma}) 
- {\bomega}({\bsigma})]\rangle 
_{{\bomega};t} \ \ .
\label{eq:domegadt}
\end{equation}
In general this 
does not yet constitute a closed set of equations 
due to the appearance 
of $p_t(\bsigma)$ in the sub-shell average.  
However, we may attempt to 
find an appropriate choice of $\bomega$ 
for which closure may be attained 
either exactly or approximately.
Thus is 
done below for certain model 
systems.

\section{Specific systems}

One particularly simple example occurs 
for the case of an infinite
range 
Ising ferromagnet in a uniform external 
field; $J_{ij}={J_0/N}$, 
$\theta_{i}=\theta^{ext}$. In this case the magnetization 
$m=N^{-1}\sum_{i}\sigma_{i}$ suffices alone as a macrovariable whose 
evolution 
is deterministic and closed,
\be
\frac{d}{dt}m=\tanh(\beta(J_0 m+\theta^{ext}))-m,
\label{eq:ferromagnet}
\ee
and 
yields the usual mean field solution in the steady state limit
\be
\frac{d}{dt}m=0~~~~~\rightarrow 
~~~~~m=\tanh(\beta(J_0 m+\theta^{ext})).
\label{eq:mft}
\ee

A greater 
challenge is posed by problems with sufficient disorder and
 frustration, 
such as those given by 
\begin{equation}
J_{ij}=J_o/N+Jz_{ij}/\sqrt{N};\ 
\ \ \ \  \ \ \langle z_{ij}\rangle =0 \ \ \ \langle {z_{ij}}^2\rangle =1 
;\ \ \ \ i\neq j.
\label{eq:exchange}
\end{equation}
where $z_{ij}$ is 
a quenched random parameter. This is the case for 
two particular model 
problems of interest on which we shall
concentrate; 
the Sherrington-Kirkpatrick 
(SK) spin glass \cite{SK} and the Hopfield 
neural network \cite{Hopfield}. 
 In the SK model the $\{\sigma\}$
represent 
true magnetic spins and the 
$\{J_{ij}\}$ are chosen randomly from a 
Gaussian distribution.
In the 
Hopfield model the $\{\sigma\}$ represent states of
McCulloch-Pitts 
neurons, 
$\sigma=+1/\!-\!1$ corresponding to firing/non-firing, 
and the $\{J_{ij}\}$ 
provide for the storage and retrieval of random 
patterns $\{\xi^\mu_i 
= \pm 1\};  \ \mu=1...p=\alpha N$, via the Hebb rule 
$
J_{ij}=N^{-1}\sum^p_{\mu=1}\xi^\mu_i\xi^\mu_j.
$ 
 Concentrating for simplicity on the region of phase space within
the 
basin 
of attraction of one pattern, $\mu=1$, it is convenient 
to apply the gauge 
transformation $\sigma_i\to\sigma_i\xi_i^1, \ \  
J_{ij}\to \xi^1_i\xi^1_j 
J_{ij}, \ \ \theta_i\to\xi^1_i\theta_i$ 
to re-write this in the form of 
(\ref{eq:exchange}) with 
\begin{equation}
z_{ij}=\frac{1}{\sqrt{p}} \sum^p_{\mu>1}
\xi^1_i\xi^\mu_i\xi^1_j\xi^\mu_j 
\ \ \ \  J_o=1 \ \ \ \ J=\sqrt{\alpha} \ \ \ \ 
\langle\xi_i\rangle=0 
\ \ .
\label{eq;Hopfield}
\end{equation}
In both cases we shall take $\theta^{ext}=0$ 
for simplicity in this
note. 
It is straightforward to show that $m=N^{-1}\sum_{i}\sigma_{i}$ 
is 
insufficient for a closed macroscopic evolution for finite $J$,
although 
it does suffice if $J\rightarrow 0$ as $N \rightarrow \infty$, as is
the 
case for a Hopfield model with only one condensed pattern, 
storing only 
a less than extensive number of patterns $(\lim _{N
\rightarrow\infty}{\alpha}=0)$. 

But how many macrovariables does one need and what are they?
Before giving 
an answer to the last question which we believe to be at least very close 
to the truth, let us consider an intermediate step which is useful illustratively. 
Although our analysis applies to finite times, it is instructive to ask 
first about the long time steady state. For problems with detailed balance 
in their dynamics one knows that in the limit as $t\rightarrow\infty$ before 
$N\rightarrow\infty$ the microstate distribution takes the Boltzmann form 
$p_\infty(\bsigma) \sim \exp (-\beta H)$. This is the case in the above 
examples which have $J_{ij}=J_{ji}$, yielding
the Hamiltonian
\begin{equation}
H/N= 
-N^{-1}\sum_{i<j} J_{ij}\sigma_i\sigma_j \  
= -\frac{1}{2}J_0m^2({\bsigma}) 
- Jr({\bsigma}) + 0(N^{-1})
\label{eq:Hamiltonian}
\end{equation}
where
\be
r({\bsigma}) 
= N^{-3/2} \sum_{i<j}\sigma_iz_{ij}\sigma_j.
\label{eq:r}
\ee
Thus, as 
long as $r(\bsigma) \sim O(1)$ it cannot be ignored in the set of $\bomega$. 
It is straightforward to show that $r(\bsigma) \sim O(1)$ for both the 
SK spin glass and the Hopfield model at finite storage ratio $\alpha$. 
Thus we shall first discuss an attempt to find a non-equilibrium macrodynamics 
in terms of $m,r$, alone, and show that it provides a reasonable but imperfect 
description. We shall then go on to a more sophisticated theory in terms 
of a  generalized order function which provides a very good fit to the 
results of microscopic simulation.

\section{The simple version of 
the theory: two order parameters}

In this section we choose the minimal 
form 
\begin{equation}
{\bomega}^s({\bsigma})\equiv(\Omega_1({\bsigma}), 
\Omega_2({\bsigma})) = (m({\bsigma}), r({\bsigma})) \ \ .
\label{eq:minomega}
\end{equation}
The 
resultant $P_t[{\bomega}^s]$ does indeed satisfy a Liouville equation in 
the thermodynamic limit, yielding the deterministic flow equations
\begin{equation}
\frac{dm}{dt}=\int 
dzD_{m,r;t}(z)\tanh \beta (J_om+Jz)-m
\label{eq:dmdt}
\end{equation}
\begin{equation}
\frac{dr}{dt}=\int 
dzD_{m,r;t}(z) z \tanh \beta(J_om+Jz)-2r
\label{eq:drdt}
\end{equation}
where 
$D_{m,r;t}(z)$ is the sub-shell averaged distribution of the disorder contributions 
to the local fields
\begin{equation}
D_{m,r;t}(z)=\lim_{N\to\infty}\frac{\sum_{\bsigma}p_t({\bsigma})\delta(m-m({\bsigma}))\delta(r-r({\bsigma}))N^{-1}\sum_i\delta(z-z_i({\bsigma}))}
{\sum_{\bsigma}p_t({\bsigma})\delta(m-m({\bsigma}))\delta(r-r({\bsigma}))}
\label{eq:Dmrt}
\end{equation}
\begin{equation}
h_i({\bsigma})=J_om({\bsigma}) 
+ Jz_{i}({\bsigma}) + 0(N^{-1}) \ \ \ \ \ \ z_i({\bsigma})=N^{-1/2}\sum_jz_{ij}\sigma_j 
\ \ .
\label{eq:hz}
\end{equation} 
As yet, because of the $p_t({\bsigma})$ 
in (\ref{eq:Dmrt}) , equations (\ref{eq:dmdt}) and (\ref{eq:drdt})  are 
not closed except in the disorder-free case $J=0$.  To close the equations 
we introduce two simple ans\"atze:
(i) we assume that the evolution of 
the macrostate $(m,r)$ is self-averaging with respect to the specific microscopic 
realization of the disorder $\{z_{ij}\}$,
(ii) as far as evaluating $D(z)$ 
is concerned we assume equipartitioning of the microstate probability $p_t({\bsigma})$ 
within each $(m,r)$ shell.
The first of these ans\"atze is well borne out 
by computer simulations of the microscopic dynamics and permits averaging 
$D(z)$ over pattern choices.  The second, which is clearly true as $t\rightarrow\infty$ 
since $p_{\infty}(\bsigma)$ depends only on $m$ and $r$ but can only be 
judged {\it a posteriori} for general time, eliminates memory effects beyond 
their reflection in $m,r$ and removes explicit time-dependence from $D$. 
Together these ans\"atze give
\begin{equation}
D_{m,r;t}(z)\to D_{m,r}(z) 
= \left\langle \frac{\sum_{\bsigma}\delta(m-m({\bsigma}))
\delta(r-r({\bsigma}))N^{-1}\sum_i\delta(z-z_i({\bsigma}))}{\sum_{\bsigma}\delta(m-m({\bsigma}))\delta(r-r({\bsigma}))}
\right\rangle_{\{z_{ij}\}}
\label{eq:Dmr}
\end{equation}
where 
$\left\langle\cdots\right\rangle_{{z_{ij}}}$ indicates an average over 
the quenched randomness. This yields closure of (\ref{eq:dmdt}) and (\ref{eq:drdt}) 
since $D_{m,r}(z)$ now depends only upon the instantaneous values of $m, 
r$ and no longer on other microscopic measures of history. 

The actual 
evaluation of $D_{m,r}(z)$ from (\ref{eq:Dmr}) remains a non-trivial exercise, 
but one which is amenable to solution by replica theory as developed for 
the investigation of local field distributions in spin glasses \cite{Thomsen}. 
 After several manipulations it can be expressed in the form
\begin{equation}
D_{m,r}(z)=\lim_{n\to 
0}\int\prod_{i,j}\prod_{\alpha,\beta=1...n}dx^\alpha_idy_j^{\alpha\beta}\exp[-N\Phi(m,r,z;\{x_i^\alpha\},\{y_j^{\alpha\beta}\})]
\label{eq:Dext}
\end{equation}
where 
the number of indices $i,j$ is finite and $\Phi$ is $O(N^0)$.  Because 
the argument of the exponential scales as $N$, the integral can be evaluated 
by steepest descents.  

The extremization is complicated \cite{SK,Thomsen,Mezard} 
and in its complete form involves significant subtleties, including an 
extension of those devised by Parisi for the analysis of the spin glass 
problem \cite{Parisi,Parisi2}. It is discussed in detail elsewhere  \cite{Coolen2,Coolen94}; 
here we note only a few salient results. Important among them is that in 
the steady state limit of $dm/dt = dr/dt =0$ the analysis yields the full 
thermodynamic results obtained from equilibrium analysis \cite{SK,Amit,Mezard}, 
including replica-symmetry breaking \cite{Parisi}. For more general times 
explicit analysis to date has only been completed within the further ansatz 
of replica-symmetry in the dynamic analogue of the spin glass order parameter 
$q^{\alpha\beta}$ which enters into the evaluation of  $D_{m,r}(z)$, but 
including a determination of the limit of its applicability against small 
replica-symmetry breaking fluctuations (cf. \cite{AT}. 

The full analytic 
results for this case can be found in \cite{Coolen2,Coolen94}. Here we 
simply exhibit graphically the comparisons between theory and microscopic 
simulation for the Hopfield model for $\alpha=0.1$ and deterministic microdynamics; 
similar results hold for arbitrary $\alpha$ and $T$. Fig 1 shows flows 
in $(m,r)$, with time implicit, and it is observed that the comparison 
is quite good (but not perfect); it also clearly shows the need for (at 
least) two order parameters. On the other hand, Fig 2, which shows the 
dependence of $m$ and $r$ on $t$, demonstrates that the theory misses a 
slowing-down effect seen in the simulations for non-retrieving situations 
(ie. ones in which $\lim_{t\rightarrow\infty}m(t)=0)$. One may further 
note that the slowing-down occurs before the system crosses the limit of 
replica-symmetry stability against small fluctuations, suggesting that 
its origin lies other than in the breakdown of the RS ansatz used in the 
evaluation of $D(z)$. R!
 ather, one is driven to conclude that the problem 
lies in the loss of memory information inherent in the assumption of equipartitioning 
in the form used above. This implies that the set of $\bomega$ must be 
expanded beyond just $m$ and $r$, to include more microscopic effects.

\begin{figure}[t]
\centering
\vspace*{83mm}
\hbox to
\hsize{\hspace*{14mm}\includegraphics{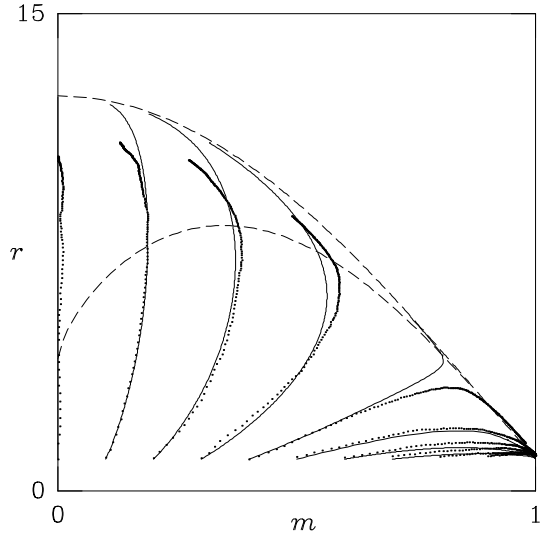}\hspace*{-14mm}}
\vspace*{-7mm}
\caption{Macroscopic 
flow trajectories for a Hopfield model with storage capacity $\alpha=0.1$ 
and deterministic microscopic dynamics $(\beta = \infty)$; dots indicate 
simulations $(N=32000)$, solid lines indicate analytic RS theory.  The 
outer dashed line is the boundary predicted by RS theory; the inner dashed 
line indicates the onset of instability against RS-breaking fluctuations, 
with stability on the side closer to the origin.}
\end{figure}
\begin{figure}[b] 

\vspace*{80mm}
\hbox to
\hsize{\hspace*{18mm}\includegraphics{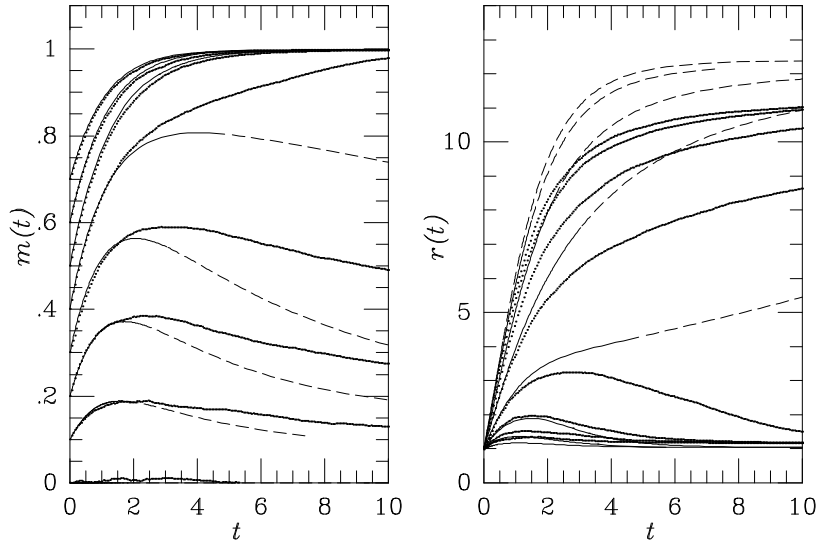}\hspace*{-18mm}}
\vspace*{-2mm}
\caption{Temporal 
dependence of the order parameters for a Hopfield model with storage $\alpha=0.1$ 
and zero-temperature dynamics $(\beta=\infty)$; dots indicate simulations 
$(N=32000)$, the other lines indicate RS theory shown with solid lines 
where stable, dashed lines where unstable.  Time is measured in Monte Carlo 
steps per spin.}
\end{figure}

\section{The sophisticated version of the 
theory: order function dynamics}

To improve on the theory as developed 
in the last section requires broadening the range of order parameters. 
Addition of a finite number of extra observables, although technically 
relatively straightforward to effectuate, is not however expected to give 
more than just minor improvements; rather, a qualitative change of philosophy 
would seem to be required. To this end we propose instead for $\bomega$ 
the joint spin-field distribution  
\begin{equation}
D(\s,h;\bsigma) 
= 
\frac{1}{N} \sum_i \delta_{\s,\sigma_i}
\delta\left[h \minus h_i (\bsigma)\right]. 
\label{eq:distribution}
\end{equation}

Our motivation for this choice 
is the following
\begin{enumerate}
\item The previous two dynamic parameters 
$m(\bsigma)$ and
$r(\bsigma)$ 
can be written
as moments off $D(\s,h;\bsigma)$, 
so the advanced theory
automatically inherits the
exactness in the two 
solvable limits $t\rightarrow\infty$ and
$J\rightarrow0$.
\item The order 
parameter function $D(\s,h)$ specifies the 
underlying states $\bsigma$ 
to a much higher degree than
$(m,r)$; i.e. more microscopic memory is taken 
into account.
\item The microscopic equation (\ref{eq:master}) itself is 
formulated
in terms of spins and fields.
\item The choice (\ref{eq:distribution}) 
allows for immediate 
generalization to models without detailed balance 
and to soft-spin models. 
\end{enumerate}   

Strictly, $D$ is infinite-dimensional 
through $h$ in the limit $N\rightarrow\infty$, but in practice we expect 
it to be quasi-continuous and well-behaved (smooth). Hence we assume that 
we can evaluate it at a number $\ell$ of field values $h_\mu$ and take 
the limit $\ell\rightarrow\infty$ after the limit $N\rightarrow\infty$. 
We then have $2\ell$ macrovariables $\Omega_{\s\mu}(\bsigma)=D(\s,h_\mu;\bsigma)$, 
with
$\mu=1,\ldots,\ell$ and $\s=\pm 1$.

We may consider the distribution
\begin{equation}
P_t[D(\s,h)]=\sum_{\bsigma} 
p_t (\bsigma) \delta[D(\s,h)-D(\s,h;\bsigma)]
\label{eq:PD}
\end{equation}
and 
analyze its evolution in a manner analagous to that applied earlier to 
$P_t(m,r)$. For finite $\ell$ it is again found to satisfy a Liouville 
equation for finite $t$ and $N\rightarrow\infty$, resulting in deterministic 
evolution of $D_t(\s,h)$. As before, this equation is not {\em{a priori}} 
closed since it involves expressions of the form
\be
\bra f (\bsigma)\ket_{D;t}=
\frac{
\sum_{\bsigma} 
p_t(\bsigma) f (\bsigma)
\prod_{\s\mu}
\delta\left[D(\s,h_\mu)\minus
\frac{1}{N}\sum_j 
\delta_{\s,\sigma_j}
\delta\left[h_\mu\minus h_j (\bsigma) \right]\right]}
{\sum_{\bsigma} 
p_t(\bsigma)  
\prod_{\s\mu}\delta \left[D(\s,h_\mu)\minus
\frac{1}{N}\sum_j 
\delta_{\s,\sigma_j}
\delta\left[h_\mu\minus h_j (\bsigma) \right]\right]}
\label{eq:subshellaverage}
\ee
where 
$p_t(\bsigma)$ again implies knowledge of the full microstate distribution. 
Again, to achieve closure, we apply our two anz\"atze of self-averaging 
and equipartitioning, but in this case the latter is far less questionable 
since it assumes only that all microstates with the same distribution $D(\s,h)$ 
contribute equally (and not all states with the same two moments of that 
distribution $m$ and $r$). After some manipulation we obtain the relatively 
simple closed evolution equation 

\bd
\frac{\partial}{\partial t} D_t(\s,h) 
= 
\frac{1}{2}\left[1\plus\s\tanh(\beta h)\right]D_t(\minus\s,h) 
-\frac{1}{2}\left[1\minus\s\tanh(\beta 
h)\right]D_t(\s,h)
\ed

\be
+\frac{\partial}{\partial h} \left\{D_t(\s,h) 
\left[h\minus\theta\minus J_0\bra\tanh(\beta
H)\ket_{D_t}\right]
+ {\cal 
A}[\s,h;D_t]
+J^2\left[1\minus\bra\sigma\tanh(\beta H)\ket_{D_t}\right]
\frac{\partial}{\partial 
h}D_t(\s,h)\right\}
\label{eq:finaldiffusion}
\ee
with
\be
{\cal A}[\s,h;D_t]=-\lim_{N\rightarrow\infty}
\frac{J}{N\sqrt{N}} 
\sum_{i\neq j}\bra\bra z_{ij}
\tanh(\beta h_i(\bsigma)) \delta_{\s,\sigma_j}
\delta[h\minus 
h_j(\bsigma)] \ket\ket_{D_t}
\label{eq:flowterm}
\ee
where
\be
\bra f(\sigma,H)\ket_D=\sum_{\sigma}\int{dH}f(\sigma,H)D(\sigma,H)
\label{eq:averageonD}
\ee
and
\be
\bra\bra 
f[\bsigma;\{z_{kl}\}]\ket\ket_{D}=\lim_{n\rightarrow 0}
\sum_{\bsigma^1}\cdots\sum_{\bsigma^n}  
\bra f[\bsigma^1;\{z_{kl}\}]\prod_{\alpha=1}^n
\prod_{\s\mu}
\delta\left[D(\s,h_\mu)\minus
\frac{1}{N}\sum_j 
\delta_{\s,\sigma^\alpha_j}
\delta\left[h_\mu\minus h_j (\bsigma^\alpha) 
\right]\right]
\ket_{\{z_{kl}\}}.
\label{eq:closedaverage}
\ee

\begin{figure}[t]
\centering
\vspace*{65mm}
\hbox 
to
\hsize{\hspace*{10mm}\includegraphics{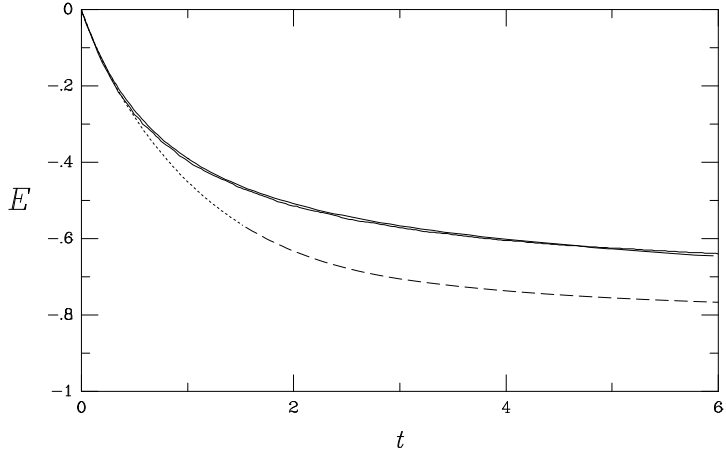}\hspace*{-10mm}}
\vspace*{-4mm}
\caption{Evolution 
of the binding energy of the
Sherrington-Kirkpatrick 
spin glass $(J_o=0)$ 
from a random microscopic start.  Comparison of 
simulations $(N=8000$, 
solid line) and predictions of the simple 
two-parameter $(m,r)$ theory 
of section IV (RS stable, dotted; RS
unstable, 
dashed) and of the advanced 
order-function theory of section V
(solid), 
for $\beta = \infty$.  Note 
that the two solid lines are almost coincident.}
\end{figure}

\begin{figure}[b]
\vspace*{98mm}
\hbox 
to
\hsize{\hspace*{-0cm}\includegraphics{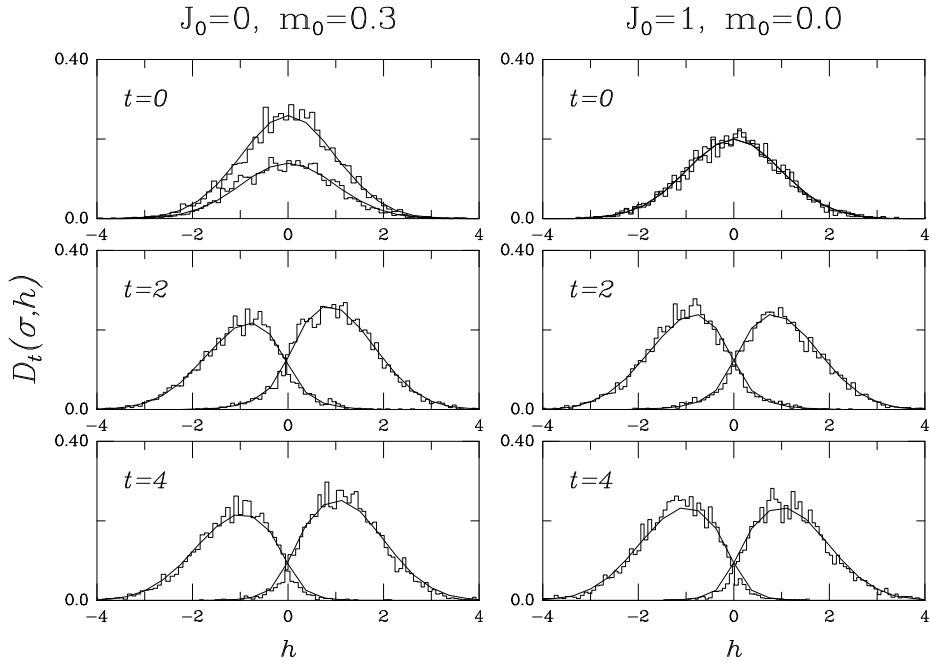}\hspace*{0cm}}
\vspace*{-2mm}
\caption{Field 
distribitions $D_t(\sigma,h)$ in the $J=1$ SK model at $T=0$, for $J_0=0$ 
(left) and $J_0=1$ (right) . Histograms:
numerical simulations with $N=8000$; 
lines: result of solving
the RS diffusion equation.}
\label{fig:distT0}
\end{figure}

Replica 
theory can again be employed to determine the averages 
(\ref{eq:closedaverage}) 
to yield a closed autonomous equation for $D_t(\s,h)$, although explicit 
numerical solution still involves the solution of a set of complicated 
saddle point equations; for details see \cite{LCS}. To date we have only 
performed the evaluation within the RS ansatz. Results for the case of 
an SK model with $J_0=0$ are exhibited graphically in Figs 3 and 4, together 
of microscopic simulation and, in Fig 3 of the simpler theory of Section 
IV. It may be noted that the advanced theory is in good accord with the 
simulations, quite convincingly describing the transients of the simulation 
experiment, including the hitherto unexplained slowing down. Equally good 
fits for $D_t(\sigma,h)$ are obtained at finite temperature (\cite{LCS}).

Replica-symmetry breaking analysis remains to be performed for arbitrary 
$t$, but we note that the difference in the asymptotic limit $t\rightarrow\infty$ 
is known from equilibrium theory where replica-symmetric theory gives $r_{RS}(\infty)=0.798$, 
while the full replica-symmetry breaking theory gives $r_{RSB}(\infty)=0.763$, 
the difference between which is very small on the scale of Fig 3.

\section{Beyond 
detailed balance}

The sophisticated version of our theory, as discussed 
in the last
section, does not make use of detailed balance and in (\ref{eq:hz}) 
$h_i(\bsigma)$ is simply given by (\ref{eq:field}), irrespective of the 
relationship between $J_{ij}$ and $J_{ji}$. The analysis goes through as 
in (\ref{eq:distribution}) to (\ref{eq:closedaverage}) even for more general 
choices of $z_{ij}$ not necessarily equal to $z_{ji}$. To illustrate its 
efficacy we have compared calculations and simulations for the case of 
an antisymmetric SK model with $z_{ij}=-z_{ji}$ with the $z_{ij};j\neq{i}$ 
independently Gaussian distributed,  with good agreement both at zero and 
finite temperature; we illustrate the results for $T=0$ dynamics in Figs 
5 and 6.

\begin{figure}[t]
\vspace*{90mm}
\hbox to
\hsize{\hspace*{-0cm}\includegraphics{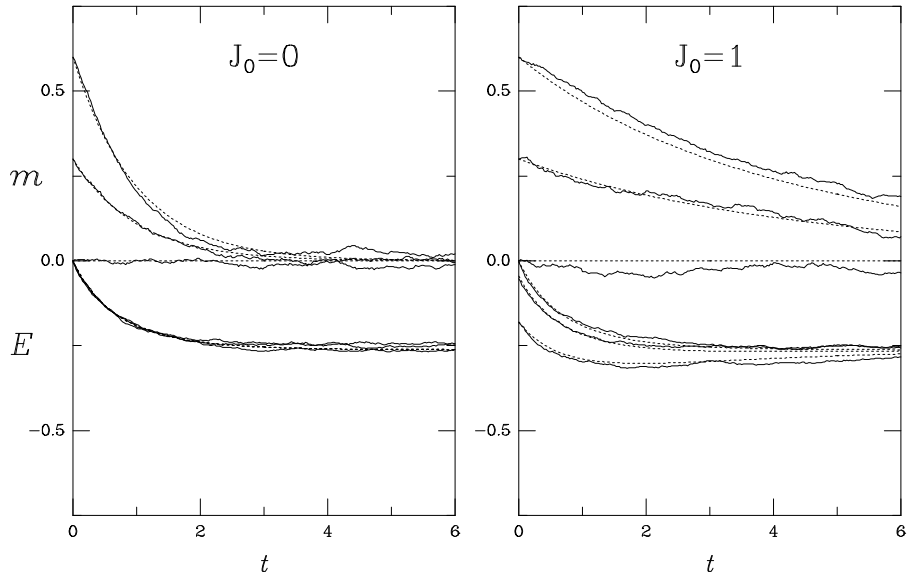}\hspace*{0cm}}
\vspace*{-2mm}
\caption{Magnetization 
$m$ and energy
per spin $E$ in the asymmetric $J=1$ SK model at $T=0$, 
for $J_0=0$ (left) and $J_0=1$ (right) . Solid lines:
numerical simulations 
with $N=5600$; dotted lines: result of solving
the RS diffusion equation.}
\label{fig:aflowT0}
\end{figure}

\begin{figure}[b]
\vspace*{90mm}
\hbox 
to
\hsize{\hspace*{-0cm}\includegraphics{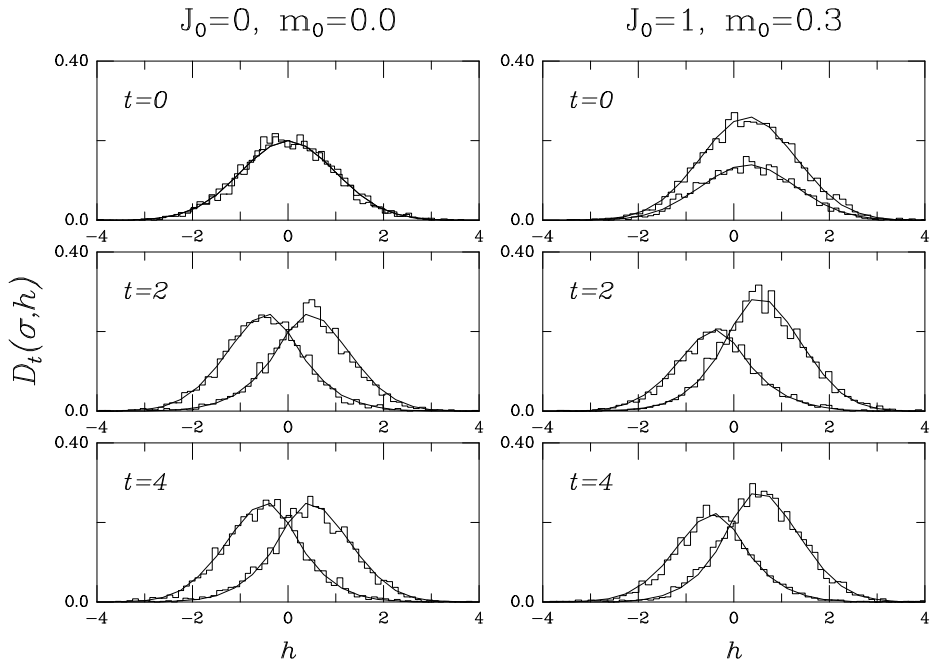}\hspace*{0cm}}
\vspace*{-2mm}
\caption{Field 
distribitions $D_t(\sigma,h)$ in the asymmetric $J=1$ SK model
at $T=0$, 
for $J_0=0$ (left) and $J_0=1$ (right) . Histograms:
numerical simulations 
with $N=5600$; lines: result of solving
the RS diffusion equation.}
\label{fig:adistT0}
\end{figure}

\section{Generalizations}

So 
far we have considered only binary-valued microvariables. However, the 
treatment above can be extended to more general microvariables ${\vec{S_i}}$, 
where the overarrow denotes a low-dimensional vector\footnote{as opposed 
to the bold notation which indicates an $N$-dimensional vector}, governed 
by some more general master equation for $p_t(\vec{\bf{S}})$. If we continue 
to restrict to situations in which the instantaneous dynamics of $\vec{S_i}$ 
is determined, possibly stochastically, by a local force 
$\vec{F_i}(\vec{\bS})$ 
then the natural extension of the order function of section V is 
\be
\calD(\vec{\calS},\vec{\calF};\vec{\bS})= 
\frac{1}{N} \sum_{i} \delta(\vec{\calS}-\vec{S}_{i}) \delta(\vec{\calF}-\vec{F}_{i}(\bS))
\label{eq:genD}
\ee
An 
equation of motion for 
\be
\calP(\calD(\vec{\calS},\vec{\calF}))=\sum_{\vec{\bS}} 
p_{t} (\vec{\bS}) \delta[\calD(\vec{\calS},\vec{\calF})-\calD(\vec{\calS},\vec{\calF};\vec{\bS})]
\label{eq:probgenS}
\ee
may 
be generated as discussed above and in appropriate limits of time and size 
can be expected to lead again to deterministic flows for $\calD(\vec{\calS},\vec{\calF})$. 
These in turn will in general involve force-fluctuation analogues of the 
noise distribution (\ref{eq:Dmrt}) and can in turn be closed via 
self-averaging 
and equipartitioning ans\"atze. We shall not, however, consider this more 
explicitly here.

It may or may not be that a macroscopic observable of 
interest can be obtained from $\calD(\vec{\calS},\vec{\calF})$ alone. If 
this is not the case the order function must be expanded to allow for their 
evaluation. 

\section{Conclusion}

We have demonstrated that for problems 
with instantaneous microscopic update dynamics, which may invove both quenched 
and stochastic randomness provided 
that the interactions are drawn from 
range-free distributions, one can derive closed autonomous deterministic 
dynamics for appropriately chosen order parameters, in good accord with 
simulations for examples studied. In principle, extensions to finite-range 
interactions could be considered but would involve greater approximation.

\section{references}


\begin{thebibliography}{Hopfield}

\bibitem{SK} 
D. Sherrington and S. Kirkpatrick, {\it Phys. Rev. Lett.} {\bf 35} (1975) 
1792
\bibitem{Hopfield} J.J. Hopfield, {\it Proc. Nat. Acad. Sci. USA} 
{\bf 79} (1982) 2554
\bibitem{Thomsen} M. Thomsen, M.F. Thorpe, T.C. Choy, 
D. Sherrington and H-J. Sommers, {\it Phys. Rev} {\bf B33} (1986) 1931
\bibitem{Mezard} 
M. M\'ezard, G. Parisi and M. Virasoro, {\it Spin Glass Theory and Beyond} 
(World Scientific, Singapore 1987).
\bibitem{Parisi} G. Parisi, {\it Phys. 
Rev. Lett.} {\bf 43} (1979) 1754
\bibitem{Parisi2} G. Parisi, {\it J. Phys} 
{\bf A13} (1980) 1101
\bibitem{Coolen2} A.C.C. Coolen and D. Sherrington, 
{\it Phys. Rev.} {\bf E49} (1994) 1921
\bibitem{Coolen94} A.C.C. Coolen 
and D. Sherrington, {\it J. Phys} {\bf A27} (1994) 7687
\bibitem{Amit} 
D.J. Amit, H. Gutfreund and H. Sompolinsky, {\it Phys. Rev. Lett.} {\bf 
55} (1985) 1530
\bibitem{AT} J.R.L. de Almeida and D.J. Thouless, {\it 
J. Phys} {\bf A11} (1978) 983
\bibitem{LCS} S.N. Laughton, A.C.C. Coolen 
and D. Sherrington, {\it J.Phys} {\bf A29} (1996) 763
\bibitem{CF} A.C.C. 
Coolen and S. Franz, {\it J. Phys} {\bf A27} (1994) 6947

\end{thebibliography}
\end{document}